# High-resolution resonant inelastic X-ray scattering study of W-L$_3$ edge in WSi$_2$[*]

ZHAO Zheqian[1], WANG Shuxing[2], WANG Xiyuan[1], SU Yang[1],
MA Ziru[3], HUANG Xinchao[4], ZHU Linfan[1]

1. Hefei National Research Center for Physical Sciences at the Microscale, Department of Modern Physics, University of Science and Technology of China, Hefei 230026, China

2. I. Physikalisches Institut, Justus-Liebig-Universität Gießen and Helmholtz Forschungsakademie Hessen für FAIR (HFHF), Campus Gießen, GSI Helmholtzzentrum für Schwerionenforschung, Gießen 35392, Germany

3. Institute of Nuclear Physics and Chemistry, China Academy of Engineering Physics, Mianyang 621999, China

4. FXE instrument, European XFEL, Schenefeld 22869, Germany

**Abstract**

With the advancement of synchrotron radiation and free-electron laser, X-ray quantum optics has emerged as a novel frontier for exploring light-matter interactions at high photon energies. A significant challenge in this field is achieving well-defined two-level systems through atomic inner-shell transitions, which are often hindered by broad natural linewidths and local electronic structure effects. This study aims to explore the potential of tungsten disilicide (WSi$_2$) as a two-level system for X-ray quantum optics applications. Utilizing high-resolution resonant inelastic X-ray scattering (RIXS) near the W-L$_3$ edge, in this work, the white line of bulk WSi$_2$ is experimentally distinguished, overcoming the spectral broadening caused by short core-hole lifetime. The measurements are conducted by using a von Hamos spectrometer at the GALAXIES beamline of the SOLEIL synchrotron. The results reveal a single resonant emission feature with a fixed energy transfer, confirming the presence of a discrete $2p$-$5d$ transition characteristic of a two-level system. Additional high-resolution XAS spectra, obtained via high energy resolution fluorescence detection method and reconstructed from off-resonant emission (free from self-absorption effect for bulk WSi$_2$ sample) method, further support the identification of a sharp white line. These findings demonstrate the feasibility of using WSi$_2$ as a model system in X-ray cavity quantum optics and establish RIXS as a powerful technique to resolve fine inner-shell structures.

# 1. Introduction

With the continuous progress of X-ray light source, X-ray quantum optics has become a new frontier research field[1–3]. The development of this field benefits from the high-brightness, high-monochromaticity and high-coherence photon source[4] provided by the new generation synchrotron radiations and X-ray free electron lasers, as well as the high-resolution and high-sensitivity detection methods developed on the basis of the X-ray sources[5,6] and the improvement of sample preparations. In recent years, the experimental and theoretical boundaries of X-ray quantum optics have been continuously broadened, and many well-known phenomena of quantum optics have been demonstrated and reproduced. In general, similar to the field of quantum optics in the visible and microwave regimes, the type of research in X-ray quantum optics strongly depends on the characteristics of the light source. The photon degeneracy in a single pulse of X-ray free electron laser (XFEL) is higher than that of synchrotron radiation, so the related research focuses on multi-photon and nonlinear phenomena, such as inner-shell lasing[7], stimulated Raman scattering[8], Rabi oscillation[9], etc. On the other hand, XFEL has unavoidable sample damage and stability problems, so the types of samples used are relatively simple, generally gas targets (soft X-ray) or thin metal foils (hard X-ray). Relatively, X-ray quantum optics research using synchrotron radiation focuses on the linear category, making full use of its richer characterization techniques and stable measurement conditions. An important branch of X-ray quantum optics, X-ray cavity quantum optics, has been derived from the combination of X-ray cavity and high-precision nano-sample preparation technology. In recent years, many quantum optical phenomena such as superradiance[10], electromagnetically induced transparency[11] and strong coupling[12] have been realized by using X-ray cavities, and abundant theoretical tools such as phenomenological quantum optical model[10,13] based on Jaynes-Cummings framework, multi-mode theory[14,15] and quantum Green's function method[15–17] have been developed. These studies not only open up new prospects for the application of quantum optics in the X-ray regime but also provide a new direction for the development of X-ray technology.

The interaction between photon and matter is the core of quantum optics. In the X-ray regime, the atomic inner shell and the nuclear energy level systems are commonly investigated. Among them, the excited state of the nucleus has a longer coherence time, which is easier to compare with the research systems in the visible region, so the initial research works mostly use the nuclear transition as the basic quantum system[18]. However, limited experimental methods restrict the study of nuclear systems[3]. Different from the atomic nucleus, the energy levels of the atomic inner shell are extremely rich, and their transition matches well with the photon energies provided by the synchrotron radiation and free electron laser, which can provide more abundant detection methods and further broaden the physical boundary of X-ray quantum optics. In recent studies, important physical phenomena, such as Fano interference[19,20] and core-hole lifetime control[21], have been realized by using dipole-allowed transitions near the $L_3$ edge of Ta and W. In addition, the directional radiation was realized by the K-edge fluorescence of Fe, Mo, Co and other elements[22]. These works show that the inner-shell system is also an outstanding platform[23] of X-ray quantum optics, and also provide a reference for exploring new phenomena.

Despite the remarkable progress in the study of quantum optics of atomic inner shells, there are still many problems to be solved. For example, the extremely short core-hole lifetime leads to serious energy broadening, and conventional experimental methods such as X-ray absorption spectroscopy (XAS) and X-ray reflectivity (XRR)[24] cannot effectively distinguish the fine inner shell transitions. In practical research, the inner shell system in X-ray regime is generally high $Z$ elements, and the sample is generally thin film. In solid materials, the energy level structure of the inner shell is usually affected by the crystal field, ligand field splitting effect and etc[25]. For example, in metal oxides such as $WO_3$[26], $V_2O_5$[27], $CeO_2$[28], the corresponding inner shell transition contains multiple features due to the possible fine structure of the excited states. Therefore, in order to accurately detect and control the excited state structure, more advanced spectroscopic techniques are needed. In recent years, resonant X-ray emission spectroscopy (RXES), also known as core to core resonant inelastic X-ray scattering (RIXS)[29], has been developed to overcome the problem of line broadening caused by core-hole lifetime. In particular, RXES can identify the fine structure with high accuracy in the measurement of the resonant state structure of the $5d$ orbital. The specific experimental methods of RXES include scanning the incident beam with a fixed energy loss, obtaining non-resonant fluorescence spectroscopy (high energy resolution off spectroscopy, HEROS) by using the incident photon energy far below the absorption edge, or obtaining high-resolution fluorescence detection absorption spectroscopy (high resolution fluorescence detection absorption spectroscopy, HERFD-XAS) by fixing a narrow fluorescence energy range. Taking the commonly used X-ray quantum optical inner-shell system Ta as an example, previous RXES studies have shown that its crystal field splitting effect is weak. As a result, the inner-shell transition appears as an emission peak[30] with a fixed energy transfer in the RXES two-dimensional map with a resolution of about 0.5 eV, which is a well-defined two-level system. For $WSi_2$, Pan et al.[31] showed that its $L_3$ transition showed well-defined two-level transition characteristics through theoretical calculation, but there is no report on its resonance emission experiment.

In this article, the $2p$ to $3d$ resonant inelastic scattering spectra ($2p3d$-RIXS) of W $L_3$ edge in $WSi_2$ were measured by using a dispersive von Hamos spectrometer[32]. The resonant inelastic emission line and the $L\alpha_1$ fluorescence line beyond the ionization threshold were accurately identified. Only a single stripe with a fixed energy transfer is observed in the experimental RXES two-dimensional map, which confirms that the W-$L_3$ edge white line of the $WSi_2$ sample exhibits a single transition characteristic. In addition, the HERFD-XAS spectrum was extracted from the RXES two-dimensional map, and the X-ray absorption spectrum (XAS) was reconstructed by using the Kramers-Heisenberg equation and the emission spectrum below the resonance energy (HEROS), which showed the characteristics of a single resonance peak.

The combination of RIXS and X-ray cavity quantum optics can not only enrich the research methods of cavity effect, but also promote the emergence of new X-ray spectroscopy techniques. Through these explorations, X-ray cavity quantum optics is expected to open up broader applications.

## 2. Theoretical background

Resonant transitions close to the absorption edge are common in the inner shell system in the X-ray regime, such as the dipole-forbidden pre-edge or the dipole-allowed white line structures. The white line, for example, was first discovered in absorption spectroscopy in 1935[33]. The fluorescence emission under this resonance condition is called as resonance emission spectrum, which shows different characteristics from the non-resonance fluorescence spectrum. For example, in 1981, Briand[34] found a white line structure of strong transition near the Mn K-edge of $KMnO_4$, and the strong white line transition caused the appearance of a new characteristic peak in the detected fluorescence emission spectrum. In 1982, Tulkki and Åberg[35] used the generalized Kramers-Heisenberg equation to describe the emission spectrum of a strong resonant white line from the perspective of resonant scattering. There is a strong resonance white line from $2p$ to $5d$ near the $L_3$ edge of W, and the RIXS process of the resonance white line excitation to produce fluorescence is shown in Fig. 1(a), and its differential cross section is given by the Kramers-Heisenberg equation:

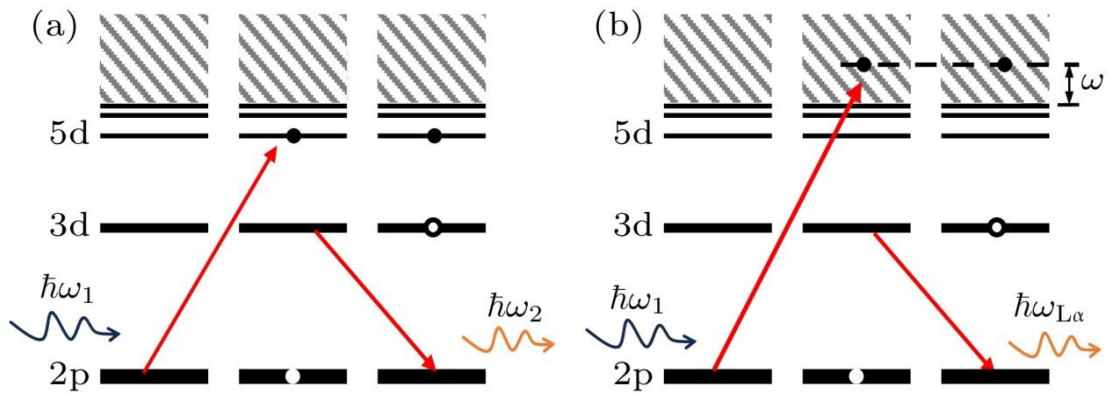

**Figure 1.** Schematic illustration of the W atom $2p$-$3d$ resonant inelastic X-ray scattering (RIXS) process with its initial, intermediate and final states shown from left to right: (a) Via the $2p^{-1}5d$ intermediate state, during which absorption and scattering are inseparable; (b) via the $2p^{-1}\epsilon d$ intermediate state, manifested as a two-step process of absorption and emission.

$$\frac{d\sigma(\omega_1)}{d\omega_2} = 2\pi r_0^2 \frac{\omega_2}{\omega_1} \frac{(\omega_{2p}-\omega_{3d})g_{3d2p}(\omega_{2p}-\omega_{5d})g_{2p5d}}{(\omega_{2p}-\omega_{5d}-\omega_1)^2+\Gamma_{2p}^2/4\hbar^2} \qquad (1)$$
$$\times \delta(\omega_1 - \omega_{3d} + \omega_{5d} - \omega_2),$$

$$\frac{d\sigma(\omega_1)}{d\omega_2} = 2\pi r_0^2 \int_0^\infty \left(\frac{\omega_2}{\omega_1}\right) \frac{(\omega_{2p}-\omega_{3d})g_{3d2p}(\omega_{2p}+\omega)}{(\omega_{2p}+\omega-\omega_1)^2+\Gamma_{2p}^2/4\hbar^2} \qquad (2)$$
$$\times \left(\frac{dg_{2p}}{d\omega}\right)\delta(\omega_1 - \omega_{3d} - \omega - \omega_2)d\omega.$$

Equation (1) describes the resonant excitation process of $2p$ to the discrete energy level $5d$ and then emitting fluorescence by re-filling $3d$ electron to $2p$, and the equation (2) corresponds to the fluorescence emission of $2p$ after ionization to the continuum state. Where $r_0$ is the classical radius of the electron, $\hbar\omega_1$ and $\hbar\omega_2$ are the incident and scattered photon energies, respectively, $\hbar\omega$ is the kinetic energy of the ionization electron, $\hbar\omega_{2p}$ and $\Gamma_{2p}$ give the binding energy of the $2p$ electron and its core-hole linewidth, $\hbar\omega_{3d}$ and $\hbar\omega_{5d}$ represent the binding energy of $3d$ and $5d$ electrons, and $g$ represents the oscillator strength of the transition between energy levels. During our experiment, the von Hamos dispersive spectrometer was used to collect the energy range of $\hbar\omega_2$ about 100 eV in a single acquisition, and scanning $\hbar\omega_1$ was used to collect the complete two-dimensional map.

Figure 1 shows the electronic transition process of the resonant inelastic scattering. If both the spin-orbit coupling and the crystal-field splitting of the $5d$ shell are weak, the upper energy-level can be regarded as a well-defined single level, and the corresponding inner-shell transition forms a single, well-defined two-level resonance. In this case, the emission energy $\hbar\omega_2$ varies with the incident photon energy, and the energy loss value is constant. Figure 1(b) shows that the inner shell electron is ionized and subsequently emit fluorescence. At this situation, the absorption and emission are almost independent two-step processes, and the fluorescence energy is fixed, such as the L$\alpha_1$ fluorescence line measured in this paper. In more general cases, spin-orbit coupling, coordination, crystal field effects in solids lead to shifts and splitting of the $5d$ band, and the resonant transition process no longer satisfies the two-level assumption. However, such fine energy-level structure will be obscured by the core-hole lifetime broadening effect, and therefore high-resolution experiments are needed to distinguish the fine structure, such as the HEROS and HERFD-XAS methods mentioned above, which are both used to study the white line transition of the W-L$_3$ edge in WSi$_2$ in this work.

Based on equations(1) and (2), we know that the scattering intensity is determined by the product of three functions, that is, a Lorentzian function with the natural line width of $\Gamma_{2p}$; a $\delta$ function to ensure energy conservation; and the distribution of the discrete upper level state(s) and the density of the states of the continuum $dg_{2p}/d\omega$, which can be simplified to the overlap of the $5d$ discrete state of W and the ionization continuum in our system. In conventional absorption spectroscopy, the details of the unoccupied states are smeared out due to the large $\Gamma_{2p}$, which is called the core-hole lifetime broadening effect. In resonant emission spectroscopy, the bandwidth of the scattering spectrum is mainly determined by the $\Gamma_{3d}$ (much smaller than the $\Gamma_{2p}$) and the instrumental resolution, so the unoccupied states can be measured with higher energy resolution[36]. For example, when the incident photon energy is low, the unoccupied state density distribution is located in the tail section of the broad Lorentzian profile, and the shape of the $dg_{2p}/d\omega$ can be measured in a high-resolution way by scanning the emission spectrum energy[37]. In addition, the distribution of $dg_{2p}/d\omega$ can also be obtained in a high-resolution manner by fixing a narrow fluorescence energy range and scanning the incident photon energy[38]. These two methods, namely HEROS and HERFD-XAS as mentioned above, are often used to overcome the effect of core-hole lifetime broadening[39].

## 3. Experimental method

The experimental measurements were performed at the GALAXIES beamline of the SOLEIL Synchrotron in France. In the experiment, in order to achieve higher photon flux and improve energy resolution, we used a double-crystal monochromator with a resolution of about $1.2 \times 10^{-4}$. A 1-mm-thick WSi$_2$ single crystal bulk sample was used in the experiment. The tungsten emission spectrum was collected by a dispersive von Hamos spectrometer equipped with eight Si (111) analyzer crystals to enhance the detection solid angle. The analyzers were set to the fourth-order Bragg reflection at the W-L$\alpha_1$ fluorescence energy, corresponding to a Bragg angle of approximately 71 °. The fluorescence emission was dispersed and focused by the von Hamos spectrometer and subsequently recorded as two-dimensional images by using a Medipix detector. Considering the limited detector size and in order to improve the signal-to-noise ratio of a single pixel, the emission images from these eight Si (111) analyzer crystals were overlapped in both dispersion and focusing directions. The schematic diagram of the experimental setup is given in Fig. 2.

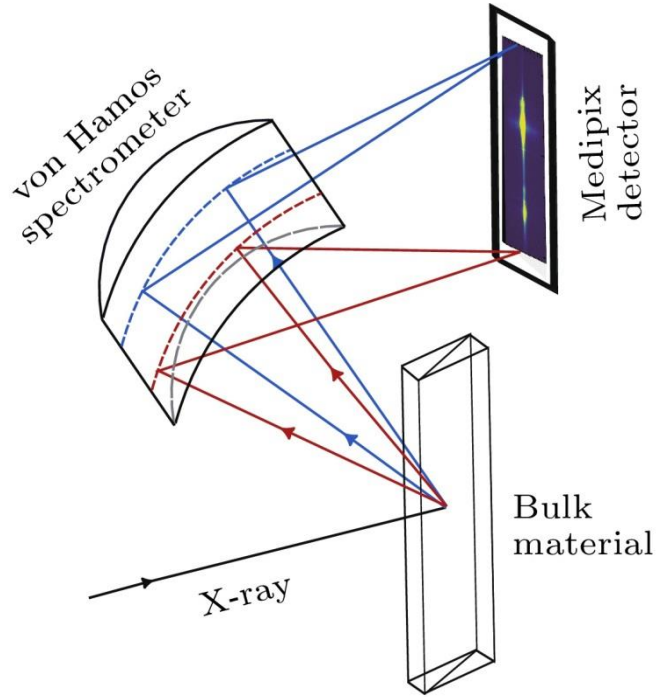

**Figure 2.** Schematic illustration of the von Hamos spectrometer, it is set to measure photon energies around 8397 eV. The energy dispersion direction is indicated by the color gradient of the X-ray beam, with red representing lower photon energies.

The W-L$\alpha_1$ emission energy of WSi$_2$ is about 8397 eV, therefore, the von Hamos spectrometer was tuned for this energy region, and correspondingly, the calibration was performed using the elastic scattering from the same sample with scanning the incident photon energy between 8310 and 8450 eV. The incident photon energy bandwidth is about 1 eV, resulted in discrete peaks on the Medipix detector. By scanning the incident photon energy and extracting the corresponding detector positions of the elastic scattering signal, a third-order polynomial fit was performed to obtain the position–energy calibration function, as shown in Fig. 3.

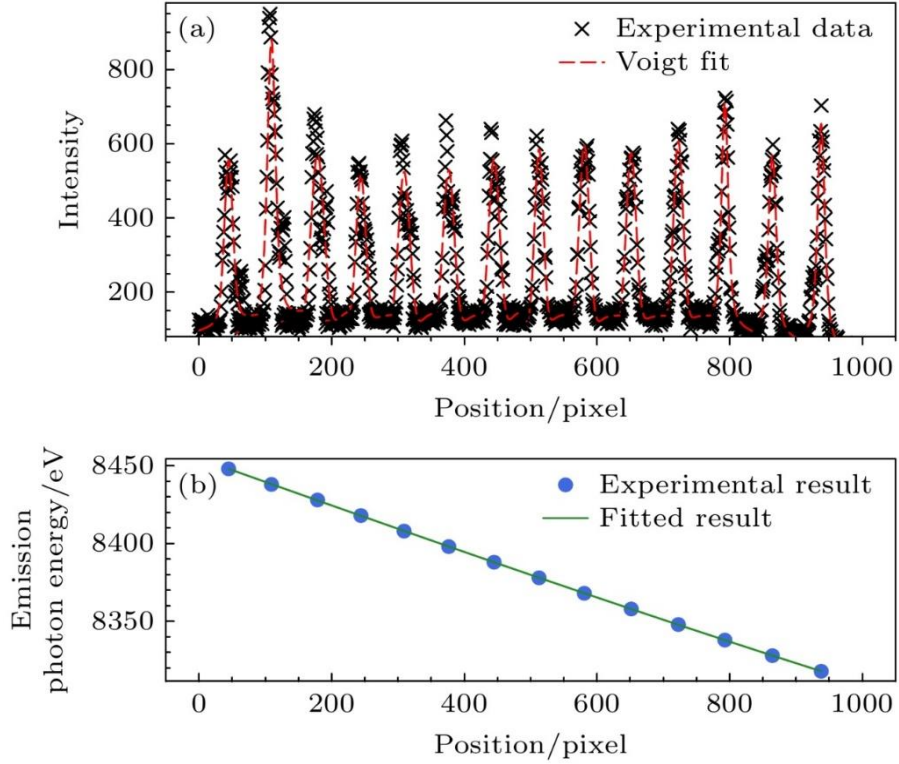

**Figure 3.** (a) Elastic scattering spectra measured at different incident photon energies; (b) each peak corresponds to a distinct position on detector along the energy-dispersive axis of the spectrometer. The dispersion relation derived by fitting the peak positions from panel (a) to panel (b), establishing the energy calibration function.

## 4. Experimental results and discussion

In the experimental measurement, the incident photon energy was scanned in the range of 10140-10250 eV, and the emission spectrum in the energy range of 8310-8450 eV was collected at each incident photon energy. In order to improve the signal-to-noise ratio, we rebin the emission spectra with 1 eV step size of incident photon energy and stack the series of emission spectra into a two-dimensional RIXS map, as shown in Fig. 4. According to Fig. 4(a), an "oblique line" (black dashed line) and a "horizontal line" (white dashed line) are observed in the 2p-5d RIXS map around the W-$L_3$ edge (~10 keV). The former corresponds to the resonant inelastic X-ray scattering process with a fixed energy transfer of $\hbar(\omega_{3d} - \omega_{5d})$, as shown in the Fig. 1(a), and the latter corresponds to the W-L$\alpha_1$ fluorescence process above ionization, as shown in the Fig. 1(b). It is obvious that the excitation below 10208 eV (energy detuning is 0) is dominated by the resonance channel transition, and the excitation above 10208 eV is dominated by the ionization channel. Figure 4(a) can also be converted into a two-dimensional spectrum with the axis of incident photon energy and energy loss, as depicted at Fig. 4(b). The emission line from the resonant inelastic scattering process in Fig. 4(b) appears as a horizontal line (black dashed line) with a fixed energy loss, while L$\alpha_1$ appear as diagonal lines (white dashed line). It is clearly shown in Fig. 4(b) that there is only one stripe with a fixed energy loss for the resonant excitation of 2p-5d, indicating that 2p-5d is a single transition.

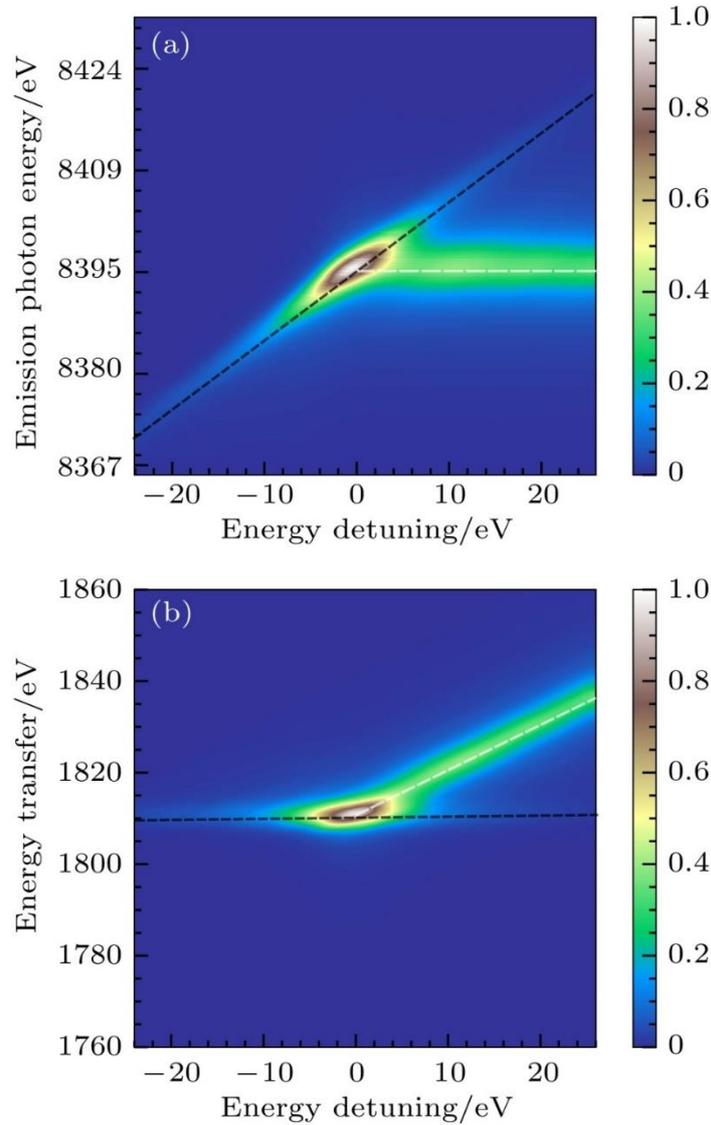

**Figure 4.** (a) Two-dimensional RIXS map near the W-L$_3$ absorption edge (10208 eV); (b) corresponding energy transfer plot. The white dashed line indicates an emission photon energy of 8397.6 eV, corresponding to the ionization of a $2p$ electron and the formation of a $2p^{-1}\epsilon d$ intermediate state. In this process, the emission energy remains constant as the incident photon energy increases, leading to a progressive increase in energy transfer. The black dashed line indicates a constant energy transfer at 1809 eV, corresponding to the $2p$-$3d$ resonant scattering process via a $2p^{-1}5d$ intermediate state, where the emission energy increases with increasing the incident energy while the energy transfer remains fixed.

The extracted RIXS spectra are shown in Fig. 5. The blue and red dashed lines indicate the $2p$-$5d$ resonant transition process and the ionization process of $2p$ electron, as labeled A and B, respectively. As shown in Fig. 5, excitation with a higher incident photon energy (10218 eV) gives rise to two peaks: the non-resonance W-L$\alpha_1$ fluorescence peak at 8397.6 eV (peak B) and a weak $2p$-$5d$ resonance fluorescence peak (peak A). With decreasing incident photon energy, peak A shifts to the left and its intensity increases gradually. At an incident photon energy of 10208 eV, peaks A and B overlap completely and become indistinguishable. As the incident photon energy is further decreased below the absorption edge of W, the L$\alpha_1$ fluorescence line disappears, leaving only the resonant transition fluorescence line (peak A).

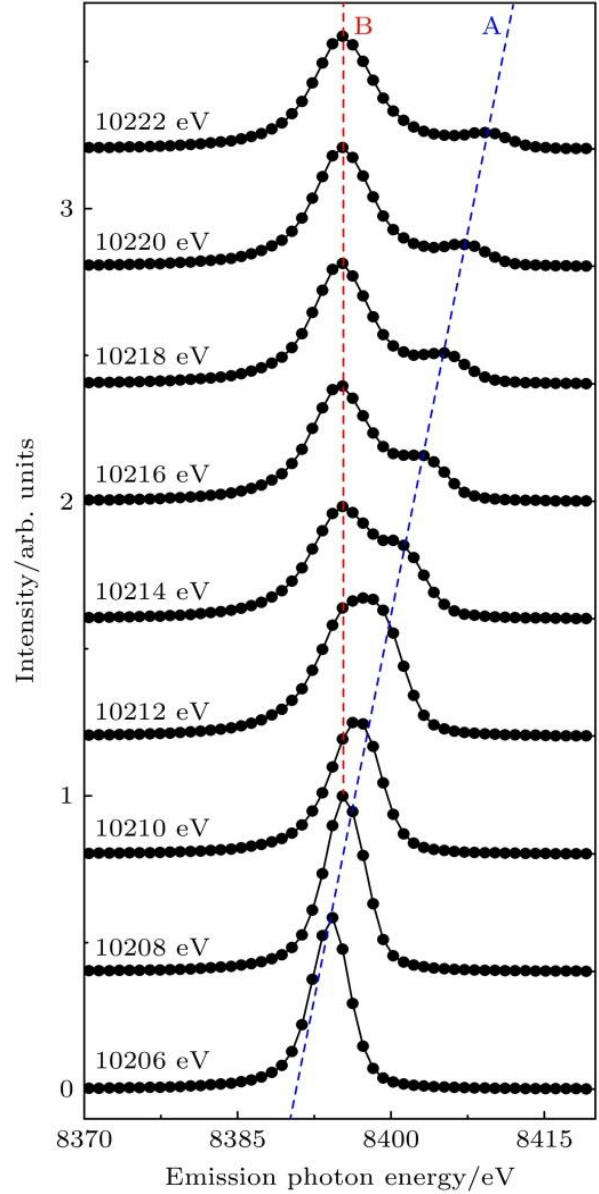

**Figure 5.** Fluorescence spectra of WSi$_2$ excited by incident photon energies from 10206 to 10222 eV, corresponding to the vertical cut shown in Fig. 4(a). The blue dashed line corresponds to the 2$p$-5$d$ resonance fluorescence peak (Peak A); the red dashed line corresponds to the non-resonance fluorescence peak of W-L$\alpha_1$ at 8397.6 eV (Peak B).

The two-dimensional RIXS map plotted in energy transfer axis is shown in Fig. 6. The blue dashed line corresponds to the 2$p$-5$d$ resonance fluorescence peak, which originates from the same feature marked by the blue dashed line in Fig. 5. When the incident photon energy increases, the W-L$\alpha_1$ non-resonance fluorescence peak appears on the higher energy transfer side.

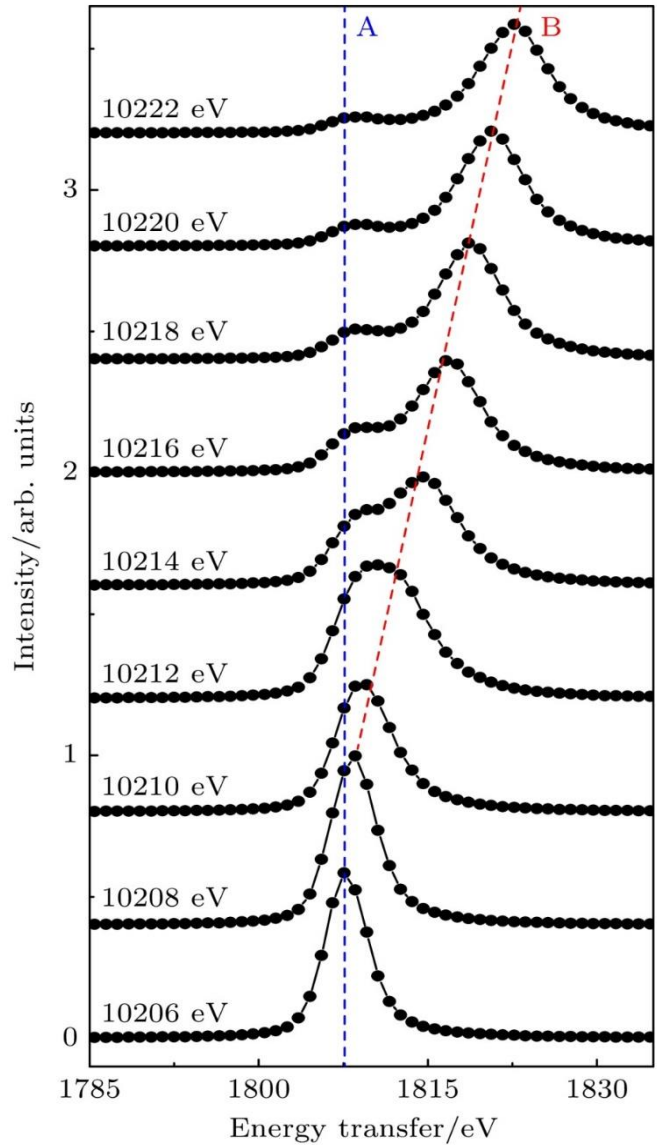

**Figure 6.** Fluorescence spectra of WSi$_2$ excited with incident photon energies from 10206 to 10222 eV, displayed as a function of energy transfer, corresponding to the vertical cut shown in Fig. 4(b). The red and blue dashed lines represent the same as those in Fig. 5.

The conventional X-ray absorption spectrum can be obtained by integrating the fluorescence over all emission energies, i.e., integrating the counts along the vertical direction in Fig. 4(a), which yields the fluorescence intensity as a function of incident photon energy (black curve of Fig. 7). Since different fluorescence channels cannot be resolved, the resonant transition and ionization channels overlap. The width of the resonance transition peak is therefore determined by the natural core-hole lifetime broadening of 2$p$. The large linewidth smears out fine spectral features, making it difficult to resolve individual transitions, as shown in the TFY curve of Fig. 7. In our previous thin-film experiments, the absorption spectrum is complex[21], and the lack of unambiguous identification of a single transition channel made the interpretation of the results challenging.

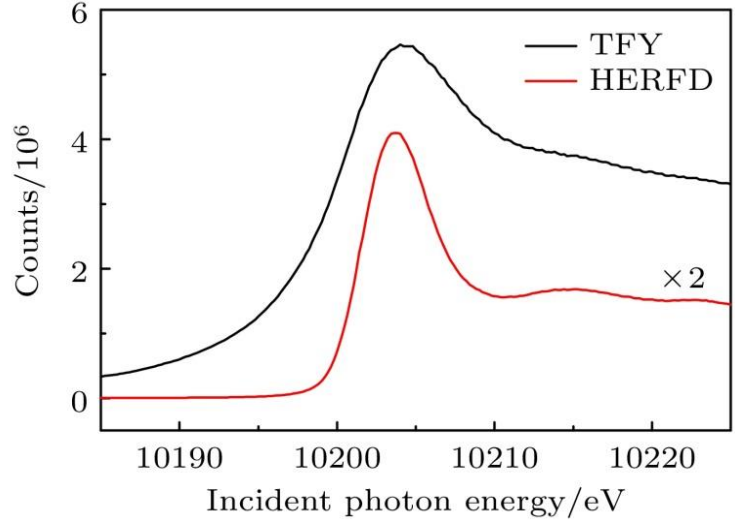

**Figure 7**. Total fluorescence yield (TFY) spectrum and high-energy resolution fluorescence detected (HERFD) spectrum in fluorescence mode. The high-resolution HERFD spectrum is obtained by integrating the X-ray emission spectroscopy (XES) data within a 0.6 eV energy window (8397–8398.4 eV) centered at the W $L\alpha_1$ fluorescence line at 8397.6 eV (indicated by the white dashed line in Fig. 4(a)). Notably, the integrated energy window is much narrower than the natural linewidth 7.2 eV of the initial state. The TFY-XAS spectrum, on the other hand, is obtained by integrating the XES intensity over the entire W $L\alpha_1$ emission range.

Unlike the total fluorescence yield (TFY) method, HERFD detects fluorescence within a narrow energy window around the $L\alpha_1$ emission line (red curve in Fig. 7). It is therefore regarded as a form of partial fluorescence yield (PFY). As evident from Fig. 4 and Fig. 7, by integrating over a narrow energy window centered at the $L\alpha_1$ emission line, HERFD yields a significantly sharper resonant peak. This enhancement effectively improves the spectral resolution and mitigates the broadening associated with the core-hole lifetime. As shown by comparing the red and the black curves in Fig. 7, the white line feature is significantly enhanced in the HERFD spectrum. In agreement with the observations from Fig. 4, the W-$L_3$ edge resonance of $WSi_2$ sample exhibits a well-defined two-level transition characteristic in HERFD spectrum. It is worth noting that neither HERFD nor TFY can avoid the self-absorption broadening effect caused by thick samples, so the line profile of HERFD in Fig. 7 is still not sharp enough.

According to equation (1), when the incident photon energy is fixed and only the emission spectrum is analyzed, the contribution of the inner-shell lifetime broadening can also be effectively suppressed. Considering the HEROS emission spectrum obtained from the incident photon energy well below the absorption edge, the scattering process can be described by generalized Kramers-Heisenberg equation:

$$I_{\text{XES}}(\hbar\omega_2) \sim \int_0^\infty \left[ \frac{\hbar\omega_2}{\hbar\omega_1} \frac{(|E_i|-|E_f|)(E+|E_i|)}{(E+|E_i|-\hbar\omega_1)^2 + \Gamma_i^2/4} \times I_{\text{XAS}}(E) \delta(\hbar\omega_1 - |E_f| - E - \hbar\omega_2) \right] dE, \quad (3)$$

The XAS spectrum can be reconstructed from the HEROS spectrum as shown in Fig. 8(a). The reconstructed XAS spectrum not only suppress the lifetime broadening problem, but also, in principle, free from self-absorption effects; Where $I_{XES}(\hbar\omega_2)$ represents the emission spectrum, $I_{XAS}(E)$ represents the XAS spectrum as function of the excited photoelectron energy $E$, $\Gamma_i$ is the initial-state broadening, $E_i$ is the energy of initial state, $E_f$ is the energy of final state, $E$ represents the energy level to where the electron is excited by the incident beam, and other symbols have the same meaning as in Eq. (1). As seen in Fig. 8(b), the absorption spectrum reconstructed from the resonant emission spectrum exhibits a significantly narrower white-line peak compared with the fluorescence yield absorption spectrum in Fig. 7. This difference originates from the influence of the fluorescence integration width and self-absorption effects in HERFD-XAS. Moreover, the white-line intensity becomes more pronounced relative to the absorption edge. The reconstructed absorption spectrum still shows a well-defined single peak feature, further confirming the two-level transition characteristics of $L_3$ edge of $WSi_2$.

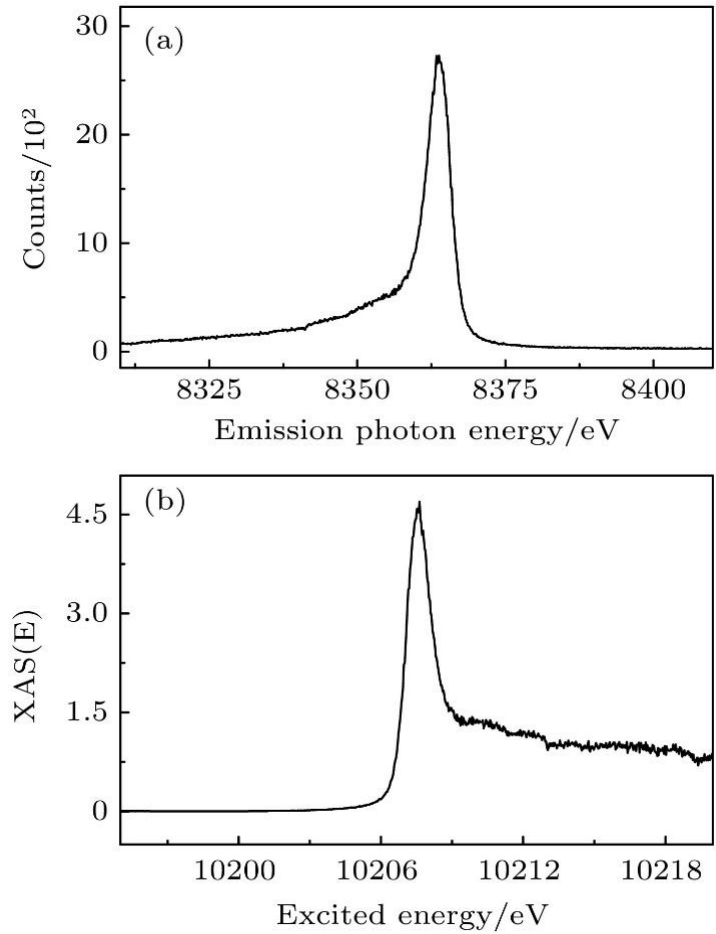

**Figure 8.** (a) Off-resonant X-ray emission spectrum (HEROS) of the sample recorded at an incident photon energy of 10172 eV; (b) reconstructed X-ray absorption spectrum (XAS) at the W-$L_3$ edge with Eq. (3).

## 5. Summary and Prospect

In this work, the resonant emission spectrum near the W-$L_3$ edge of $WSi_2$ single crystal was measured by using a high-resolution von Hamos spectrometer. In the two-dimensional RIXS map, the constant-energy-transfer feature appears as a single diagonal line, which confirms that the $2p$ to $5d$ resonance transition of $WSi_2$ is a single channel and two-level transition. We further obtained a high-resolution absorption spectrum by using the HERFD method and reconstructed the second high-resolution absorption spectrum from the HEROS spectrum, which is free from self-absorption effects. Both spectra display a single white-line feature. In addition, we demonstrate that the resonant X-ray emission spectroscopy technique can distinguish the resonant and ionization channels, providing an experimental reference for the future studies of thin-film cavities embedded with $WSi_2$ layers.

Thanks to Mr. Li Bo and other students in our group for their discussions and suggestions. We thank Dr. Yohei Uemura of FXE instrument of European XFEL, Dr. Azat Khadiev of the PETRA-III, and other collaborators for their support in the beam time. We thank Dr. James Michael Ablett and Dr. Jean-Pascal Rueff of SOLEIL synchrotron for their support in the beamtime and for providing single crystal samples.